# Spontaneous Emission in Nonlocal Materials

Pavel Ginzburg[1$=], Diane Roth[1=], Mazhar E. Nasir[1=], Paulina Segovia Olvera[1$], Alexey V. Krasavin[1=], James Levitt[1], Liisa M. Hirvonen[1], Brian Wells[2#], Klaus Suhling[1], David Richards[1], Viktor A. Podolskiy[2], Anatoly V. Zayats[1]

[1]Department of Physics, King's College London, Strand, London WC2R 2LS, United Kingdom
[2]Department of Physics and Applied Physics, University of Massachusetts Lowell, One University Ave, Lowell, MA, 01854, USA

[$] Current address: School of Electrical Engineering, Tel Aviv University, Tel Aviv 69978, Israel
[#] Current address: Dept of Physics, University of Hartford, West Hartford, CT 06117, USA
[=] equal contributions

**Abstract (100-150 words):**

Light-matter interactions can be dramatically modified by the surrounding environment. Here we report on the first experimental observation of molecular spontaneous emission *inside* a highly nonlocal metamaterial based on a plasmonic nanorod assembly. We show that the emission process is dominated not only by the topology of its local effective medium dispersion, but also by the nonlocal response of the composite, so that metamaterials with different geometric parameters but the same local effective medium properties exhibit different Purcell factors. A record-high enhancement of a decay rate is observed, in agreement with the developed quantitative description of the Purcell effect in a nonlocal medium. An engineered material nonlocality introduces an additional degree of freedom into quantum electrodynamics, enabling new applications in quantum information processing, photo-chemistry, imaging, and sensing.

**One Sentence Summary (150 characters):**

Nonlocal metamaterial properties are shown to determine a record-high broadband enhancement of decay rate of quantum emitters inside plasmonic composites.

**Main text**

When the behavior of a physical system at a given point depends on its state at another spatially separated region, the system is described as being nonlocal. Quantum states of light and matter are inherently nonlocal, reflecting the fundamental wave-particle duality [1]. Quantum entanglement is one of the most fascinating examples of nonlocalities in nature [2], [3]. Its successful demonstration in optics with photons [4], [5] is enabled by inherently weak photon-photon interactions. Material systems, on the other hand, suffer from various decoherence effects, such as electron-electron, electron-phonon and other scattering mechanisms that virtually eliminate optical nonlocality in homogeneous room-temperature media [6]. However, electromagnetic nonlocality may re-emerge in engineered composites, metamaterials [7], where coherent surface plasmons mediate coupling between unit cells of artificial electromagnetic crystals.

Here, we analyze both experimentally and theoretically the process of spontaneous emission in a nonlocal environment using a plasmonic nanorod metamaterial platform that has been recently demonstrated to enable a topological transition between elliptic and hyperbolic dispersions. Experimentally, we demonstrate a broadband, macroscopically averaged lifetime reduction of the order of 30 for several fluorophores, while microscopic reductions in lifetime are orders of magnitude higher, as estimated from the experimental data. We develop new theoretical and numerical approaches capable of calculating the local density of photonic states in nonlocal media. While nonlocal effects are known to have a weak impact on linear reflection and transmission through the metamaterial, the situation is completely different in the quantum optical regime where nonlocality results in an additional propagating mode inside the metamaterial, fundamentally altering the density of photonic states. The dynamics of emitters' decay, and other quantum optical processes, *inside* a nanorod metamaterial are essentially dominated by the nonlocal response of the composite.

*Artificial anisotropic media.* A plasmonic nanorod metamaterial (Fig.1a), a composite formed by array of aligned metallic nanorods arranged inside the dielectric host, represents an example of strongly anisotropic uniaxial media that exhibit negative permittivity (realized in reflective metals) in one direction, and positive permittivity (seen in transparent dielectrics) in other directions (Fig. 1(b)). Mathematically, the optical properties of uniaxial materials are described by a diagonal tensor of dielectric permittivity $\hat{\epsilon}$ with $\epsilon_x = \epsilon_y \equiv \epsilon_\perp \neq \epsilon_z$. Homogeneous anisotropic media support the propagation of two types of waves that differ by their polarization. Ordinary waves that have electric field normal to the metamaterial's optical axis $\vec{E} \perp \hat{z}$ do not experience an anisotropy of dielectric permittivity. On the other hand, extraordinary waves that have magnetic field $\vec{H} \perp \hat{z}$ are strongly affected by material anisotropy. The dispersion of the extraordinary waves, the relationship between components of the wavevector $\vec{k}$, angular frequency $\omega$, and speed of light $c$, is given by

$$\frac{k_x^2+k_y^2}{\epsilon_z} + \frac{k_z^2}{\epsilon_\perp} = \frac{\omega^2}{c^2} \quad \dots \dots (1)$$

The isofrequency surface which is defined as $\{k_x(\omega), k_y(\omega), k_z(\omega)\}$ forms either an ellipsoid or a hyperboloid, depending on the relationship between the signs of $\epsilon_\perp$ and $\epsilon_z$ which, in turn, are wavelength-dependent. The geometrical properties of this isofrequency surface is often referred to as the optical topology of a metamaterial [8]. The effective permittivity of the metamaterials used in our study (Fig. 1(b)) shows that the composite operates in the elliptic regime for wavelengths below 575 nm, exhibits epsilon-near-zero response at around 575 nm, and operates in the hyperbolic regime for longer wavelengths. Optical topology predicted by Eq. (1) exhibits typical ellipsoidal isofrequencies for TE modes in both elliptic and hyperbolic regimes (Fig. 1 c,d). The dispersion of TM modes in the hyperbolic regime is described by a hyperbola (Fig. 1 f). Dispersion of TM modes in the elliptic regime is strongly affected by material losses (cf. Figs. S6 g-i) and represents an ellipsoid in the centre of the Brillouin zone which is increasingly deformed for higher wavenumbers (Fig. 1 e).

The optical topology of metamaterials has a dramatic effect on their quantum optical properties, leading, theoretically, to a singularity in the local density of optical states in homogeneous hyperbolic media [9] (this effect is limited by granularity in metamaterials [10]). The density of optical states affects the rate of spontaneous emission [11], nonradiative energy transfer between molecules [12], and other processes. Recent experimental studies, mostly focused on emitters near a layered metamaterial design, have demonstrated relatively modest 5—10 fold enhancements of the decay rate in the hyperbolic dispersion case and smaller enhancements in the elliptic dispersion regime [8]. Similar results were reported for emitters on top of hyperbolic plasmonic nanorod metamaterials [13]. However, the geometry of these previous studies prevented a test of emission dynamics in the nonlocal metamaterial regime, most pronounced inside nanorod metamaterials operating in the elliptical regime [7],[14],[15],[16] (corresponding to $\lambda < 575$ nm for the metamaterials used in this work).

**Optical nonlocalities in nanorod media**. There exists a frequency range in which the nanorod metamaterial supports the propagation of not two, but three different modes [16]. One of these modes is a TE-polarized ordinary wave, while the other two waves have identical (TM) polarization, but different effective indices. The topology of these waves is shown in Fig.2(c,d). Comparing Fig.2 and Fig.1, it is seen that while in the hyperbolic regime the differences are minimal, the behavior in the elliptic regime is drastically different. The isofrequency of one of the TM polarized waves now qualitatively agrees with an ellipsoid, predicted by Eq. (1). The second, additional TM polarized mode which is absent in the local effective medium theory, has a hyperbolic-like topology (Fig 2c).

This additional wave represents a collective excitation of cylindrical surface plasmons propagating on metallic nanorods [16]. This mode is a metamaterial analog of collective light-matter excitations, exciton-polaritons, that are known to enable additional waves in homogeneous semiconductors [6] . In both homogeneous and composite media, properties of additional waves can be successfully described by introducing nonlocality, i.e. a wave-vector dependence of the permittivity of the material. However, while the permittivity of homogeneous media is essentially fixed by nature, the electromagnetic response of composites can be engineered.

In particular, the dispersion of modes in plasmonic nanorod composites can be controlled by scaling the unit cell, a process that does not change the local effective medium response (see Supplementary Information (SI) and Ref. [16]). The predicted changes in dispersion of optical modes as a result of geometry scaling are summarized in Fig. 2 (e,f). Material losses can be controlled by the choice of plasmonic metal or by fabrication (e.g., annealing) and serve as an additional degree of freedom for engineering the dispersion of the modes in a metamaterial. SI describes the correspondence between predictions of local and nonlocal effective medium theories (EMTs) and their changes in response to geometry and loss scaling in detail. From the point of view of a quantum emitter, an additional electromagnetic wave represents a separate emission decay channel that, in turn, affects its radiation dynamics, and drastically increases the density of optical states in the elliptic regime.

***Experimental studies.*** The analysis of decay dynamics of four fluorescent dyes with different emission wavelengths spanning elliptic and hyperbolic regimes and placed inside the metamaterial is summarized in Fig. 3 (see SI for detailed analyses of the decay dynamics). While all the studied dyes show nearly single-exponential fluorescence decay in a homogeneous environment, the presence of plasmonic surfaces and nanostructures makes the dynamics multi-exponential. This effect does not represent the strong coupling regime of interaction [17], but rather is the direct result of many emitters contributing to the signal, with each individual emitter having its own position- and polarization-dependent decay rate in the structured environment.

Here we have used the inverse Laplace transform [18] which does not rely on any preliminary assumptions and provides the distribution of the lifetimes present in the observations. Dye molecules situated above a glass slide show smooth localized lifetime distributions peaked around their standard values of a few nanoseconds (Figs. S3). The comparison of emission for glass and Au interfaces reveals the expected behavior, caused by the presence of the metal, resulting in excitation of surface plasmon polariton waves [19]. Once the dependence of the lifetime on the emitter's spatial position is taken into account (see SI), an almost perfect agreement between theoretical predictions and experimental measurements of lifetime distributions is achieved for the emitters near the metal interface (Fig. S3). For example, the lifetime distribution of Fluorescein (Fig. S3 (b)) is peaked at 3.9 ns, representing the reduction of the macroscopically averaged lifetime by a factor of approximately 1.2 compared to emission in vacuum. Spatial averaging takes into account the random distribution of molecules inside the sample volume. Thus, various possible collective effects which could affect the dynamics, such as super-fluorescence [20], [21], can be then ruled out.

The dynamics measured for the emitters inside the metamaterial exhibits even faster decay with an even broader distribution (7—100 times) of the lifetimes than for a single metal interface (Figs. 3e and S3). The span of the lifetime distributions reflects a strong position-dependent decay rate enhancement for the emitters inside the nanorod array unit cell [22]. Substantial lifetime reduction was observed for all wavelengths, possibly limited by the instrument function of the measurements.

The spectral dependence of the enhancement of the emission rates observed in experiments and obtained from direct numerical solutions of Maxwell equations have similar features of a broad spread of the lifetimes at all emission wavelengths and weak wavelength dependence of this distribution (Fig. 3b). The experimental data show that the distribution of decay rate enhancements, from 10 to 100 (for the high and low-lifetime cut-offs, respectively, at 10% of the modal lifetime contribution), is almost independent of the wavelength across the elliptic, epsilon-near-zero and hyperbolic regimes (Fig. 3e). Local EMT predicts a strong enhancement of the decay rate in the epsilon-near-zero regime, in sharp contrast with the observed behavior. At the same time, the experimental data are in closer agreement with predicted wavelength dependence of the decay rate given by nonlocal EMT. The remaining quantitative difference, between the predictions of nonlocal effective medium theory (that averages the response of the composite) and experimental data, stems from strong dependence of density of optical states on the position of the dipole inside the unit cell (Fig. S8). Direct numerical solutions of the Maxwell equations in the nanorod composites (see Fig. S8 and SI for details) confirm this position dependence and are also in agreement with the experimental data and the nonlocal theory. It should be noted that neither local nor nonlocal effective medium calculations allow discrimination of the emitter position in the cell of the array. Therefore, both EMTs provide an estimate of the effective average lifetime. Nevertheless, the transfer-matrix-based mode-matching formalism (as described in SI) allows further averaging of this effective lifetime over emitter position along the nanorod length (Figs. 3b,c).

It is important to underline that both local and nonlocal effective medium theories predict almost identical transmission and reflection spectra of metamaterials (Fig.S5), since the plane wave, incident from vacuum, only weakly couples to high-index additional waves inside the metamaterial. At the same time, point dipoles emitting from within the composite can successfully couple to these high-index modes, enabled by spatial dispersion. In fact, emission into additional waves dominates the decay dynamics while the metamaterial operates in the elliptical regime.

***Engineering the dispersion of the additional wave.,*** Adjusting the geometry and/or composition of a metamaterial, opens the door for engineering the quantum optics inside the composites. This is illustrated in Fig.3c where the lifetime dynamics in the composites studied above are compared to dynamics of emission in a similar metamaterial with nanorod diameter and unit cell dimension scaled by 50%. It is interesting to note that the reduction of the unit cell does not make the optical response of metamaterial appear more local; i.e., it does not lead the enhancement of local density of states at epsilon-near-zero and hyperbolic frequencies. Rather, 3D numerical solutions of Maxwell equations and nonlocal effective medium theory predict that the reduction of unit cell size yields a further, order of magnitude, broadband enhancement of decay rate, despite the local effective medium parameters of the metamaterial being the same.

Note that both nonlocal EMT and numerical solutions of Maxwell equations predict comparable enhancement of decay rate as a result of the unit cell reduction. As before, the quantitative difference between predictions of nonlocal EMT and numerical solutions reflect the limitation of an effective medium response, that nevertheless reveals the important physics and provides a relatively fast estimate of the radiation decay enhancement. Numerical solutions of Maxwell equations enable one to calculate the detailed optical response at any location at the expense of computational complexity and time.

## *Conclusion*

In summary, the effect of spontaneous emission inside an artificially created highly nonlocal medium was experimentally demonstrated and theoretically analyzed. It was shown that the effect of structural nonlocality has a major impact on the quantum electrodynamics inside nanorod media. In particular, it dominates the spontaneous emission, setting fundamental limits on the Purcell enhancement. It also enables the design of the local density of optical states via geometrical parameters of the metamaterial composite, opening a new road for engineering density of optical states with (meta) materials.

**References:**

1]   L. de Broglie, "XXXV. A tentative theory of light quanta," *Philos. Mag. Ser. 6*, vol. 47, no. 278, Apr. 1924.

[2]   A. Einstein, B. Podolsky, and N. Rosen, "Can Quantum-Mechanical Description of Physical Reality Be Considered Complete?," *Phys. Rev.*, vol. 47, no. 10, pp. 777–780, May 1935.

[3]   J. S. Bell, "On the Problem of Hidden Variables in Quantum Mechanics," *Rev. Mod. Phys.*, vol. 38, no. 3, pp. 447–452, Jul. 1966.

[4]   S. J. Freedman and J. F. Clauser, "Experimental Test of Local Hidden-Variable Theories," *Phys. Rev. Lett.*, vol. 28, no. 14, pp. 938–941, Apr. 1972.

[5]   A. Aspect, P. Grangier, and G. Roger, "Experimental Realization of Einstein-Podolsky-Rosen-Bohm Gedankenexperiment : A New Violation of Bell's Inequalities," *Phys. Rev. Lett.*, vol. 49, no. 2, pp. 91–94, Jul. 1982.

[6]   V. M. Agranovich and V. L. Ginzburg, *Spatial Dispersion in Crystal Optics and the Theory of Excitons*. Pergamon Press: New York, 1984.

[7]   R. J. Pollard, A. Murphy, W. R. Hendren, P. R. Evans, R. Atkinson, G. A. Wurtz, A. V Zayats, and V. A. Podolskiy, "Optical nonlocalities and additional waves in epsilon-near-zero metamaterials.," *Phys. Rev. Lett.*, vol. 102, no. 12, p. 127405, Mar. 2009.

[8]   H. N. S. Krishnamoorthy, Z. Jacob, E. Narimanov, I. Kretzschmar, and V. M. Menon, "Topological transitions in metamaterials.," *Science*, vol. 336, no. 6078, pp. 205–9, Apr. 2012.

[9]   Z. Jacob, I. I. Smolyaninov, and E. E. Narimanov, "Broadband Purcell effect: Radiative decay engineering with metamaterials," *Appl. Phys. Lett.*, vol. 100, no. 18, p. 181105, Oct. 2012.

[10]  A. N. Poddubny, P. A. Belov, P. Ginzburg, A. V. Zayats, and Y. S. Kivshar, "Microscopic model of Purcell enhancement in hyperbolic metamaterials," *Phys. Rev. B*, vol. 86, no. 3, p. 035148, Jul. 2012.

[11]  E. M. Purcell, "Spontaneous Emission Probabilities at Radio Frequencies," *Phys. Rev.*, vol. 69, no. 11–12, pp. 674–674, Jun. 1946.

[12]  T. U. Tumkur, J. K. Kitur, C. E. Bonner, A. N. Poddubny, E. E. Narimanov, and M. A. Noginov, "Control of Förster energy transfer in the vicinity of metallic surfaces and hyperbolic metamaterials.," *Faraday Discuss.*, vol. 178, pp. 395–412, Jan. 2015.

[13]  T. Tumkur, G. Zhu, P. Black, Y. A. Barnakov, C. E. Bonner, and M. A. Noginov, "Control of spontaneous emission in a volume of functionalized hyperbolic metamaterial,"

*Appl. Phys. Lett.*, vol. 99, no. 15, p. 151115, Oct. 2011.

[14] G. A. Wurtz, R. Pollard, W. Hendren, G. P. Wiederrecht, D. J. Gosztola, V. A. Podolskiy, and A. V Zayats, "Designed ultrafast optical nonlinearity in a plasmonic nanorod metamaterial enhanced by nonlocality.," *Nat. Nanotechnol.*, vol. 6, no. 2, pp. 107–11, Feb. 2011.

[15] K.-T. Tsai, G. A. Wurtz, J.-Y. Chu, T.-Y. Cheng, H.-H. Wang, A. V Krasavin, J.-H. He, B. M. Wells, V. A. Podolskiy, J.-K. Wang, Y.-L. Wang, and A. V Zayats, "Looking into meta-atoms of plasmonic nanowire metamaterial.," *Nano Lett.*, vol. 14, no. 9, pp. 4971–6, Sep. 2014.

[16] B. M. Wells, A. V. Zayats, and V. A. Podolskiy, "Nonlocal optics of plasmonic nanowire metamaterials," *Phys. Rev. B*, vol. 89, no. 3, p. 035111, Jan. 2014.

[17] M. O. Scully and M. S. Zubairy, *Quantum Optics*. Cambridge University Press, 1997.

[18] S. W. Provencher, "Inverse problems in polymer characterization: Direct analysis of polydispersity with photon correlation spectroscopy," *Die Makromol. Chemie*, vol. 180, no. 1, pp. 201–209, Jan. 1979.

[19] G. W. Ford and W. H. Weber, "Electromagnetic effects on a molecule at a metal surface," *Surf. Sci.*, vol. 109, no. 2, pp. 451–481, Aug. 1981.

[20] R. Bonifacio and L. Lugiato, "Cooperative radiation processes in two-level systems: Superfluorescence," *Phys. Rev. A*, vol. 11, no. 5, pp. 1507–1521, May 1975.

[21] D. Martín-Cano, L. Martín-Moreno, F. J. García-Vidal, and E. Moreno, "Resonance energy transfer and superradiance mediated by plasmonic nanowaveguides.," *Nano Lett.*, vol. 10, no. 8, pp. 3129–34, Aug. 2010.

[22] A. P. Slobozhanyuk, P. Ginzburg, D. A. Powell, I. Iorsh, A. S. Shalin, P. Segovia, A. V. Krasavin, G. A. Wurtz, V. A. Podolskiy, P. A. Belov, and A. V. Zayats, "Purcell effect in hyperbolic metamaterial resonators," *Phys. Rev. B*, vol. 92, no. 19, p. 195127, Nov. 2015.

[23] J. R. Lakowicz, *Principles of Fluorescence Spectroscopy*, 3rd ed. Springer, 2006.

[24] W. Vogel and D.-G. Welsch, *Quantum Optics*. Wiley-VCH; 3rd, Revised and Extended Edition edition, 2006.

[25] R. Wangberg, J. Elser, E. E. Narimanov, and V. A. Podolskiy, "Nonmagnetic nanocomposites for optical and infrared negative-refractive-index media," *J. Opt. Soc. Am. B*, vol. 23, no. 3, p. 498, Mar. 2006.

[26] P. H. Lissberger and R. G. Nelson, "Optical properties of thin film Au-MgF2 cermets," *Thin Solid Films*, vol. 21, no. 1, pp. 159–172, Mar. 1974.

[27] S. R. J. Brueck, "Radiation from a dipole embedded in a dielectric slab," *IEEE J. Sel. Top. Quantum Electron.*, vol. 6, no. 6, pp. 899–910, Nov. 2000.

[28] V. Podolskiy, P. Ginzburg, B. Wells, and A. Zayats, "Light emission in nonlocal plasmonic metamaterials," *Faraday Discuss.*, vol. 178, pp. 61–70, Oct. 2015.

[29] A. C. Atre, B. J. M. Brenny, T. Coenen, A. García-Etxarri, A. Polman, and J. A. Dionne, "Nanoscale optical tomography with cathodoluminescence spectroscopy.," *Nat. Nanotechnol.*, vol. 10, no. 5, pp. 429–36, May 2015.

[30] L. Novotny and B. Hecht, *Principles of Nano-Optics*, 2nd ed. Cambridge University Press, 2012.

**Figures and Figure Captions**

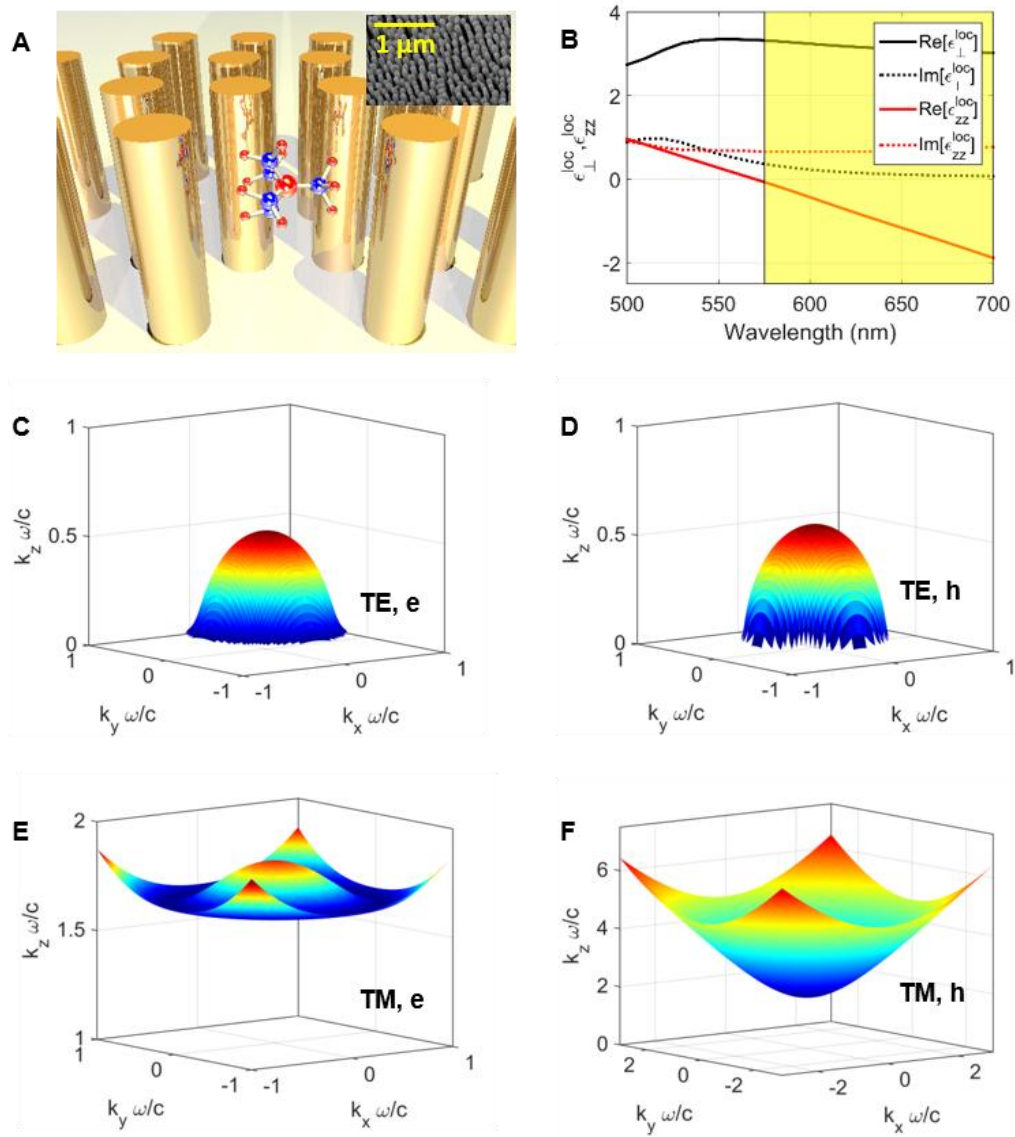

Figure 1. (A) Schematics of a nanorod metamaterial with a fluorescent molecule inside. Inset shows SEM image of the metamaterial comprised of Au nanorods in air (scale bar is 1 μm). (B) *Local* effective medium parameters of the metamaterials (Au nanorods in ethanol) determined from the geometry of the nanorod metamaterial 50±2 nm diameter, 100 nm period and 250±5 nm height) which reproduce the experimental extinction spectra (Fig. S1). (C-F) Isofrequency surfaces describing the dispersion of the propagating modes in the metamaterial shown in (A), calculated with the local EMT: (C,E) elliptic (denoted e) regime, $\lambda = 550\ nm$, (D,F) hyperbolic (denoted h) regime, $\lambda = 650$ nm for (C,D) TE- and (E,F) TM-polarized modes. The non-monotonic behavior of the TM-polarized mode in the elliptical regime (D) is a direct consequence of the material absorption (Fig. S4)
13

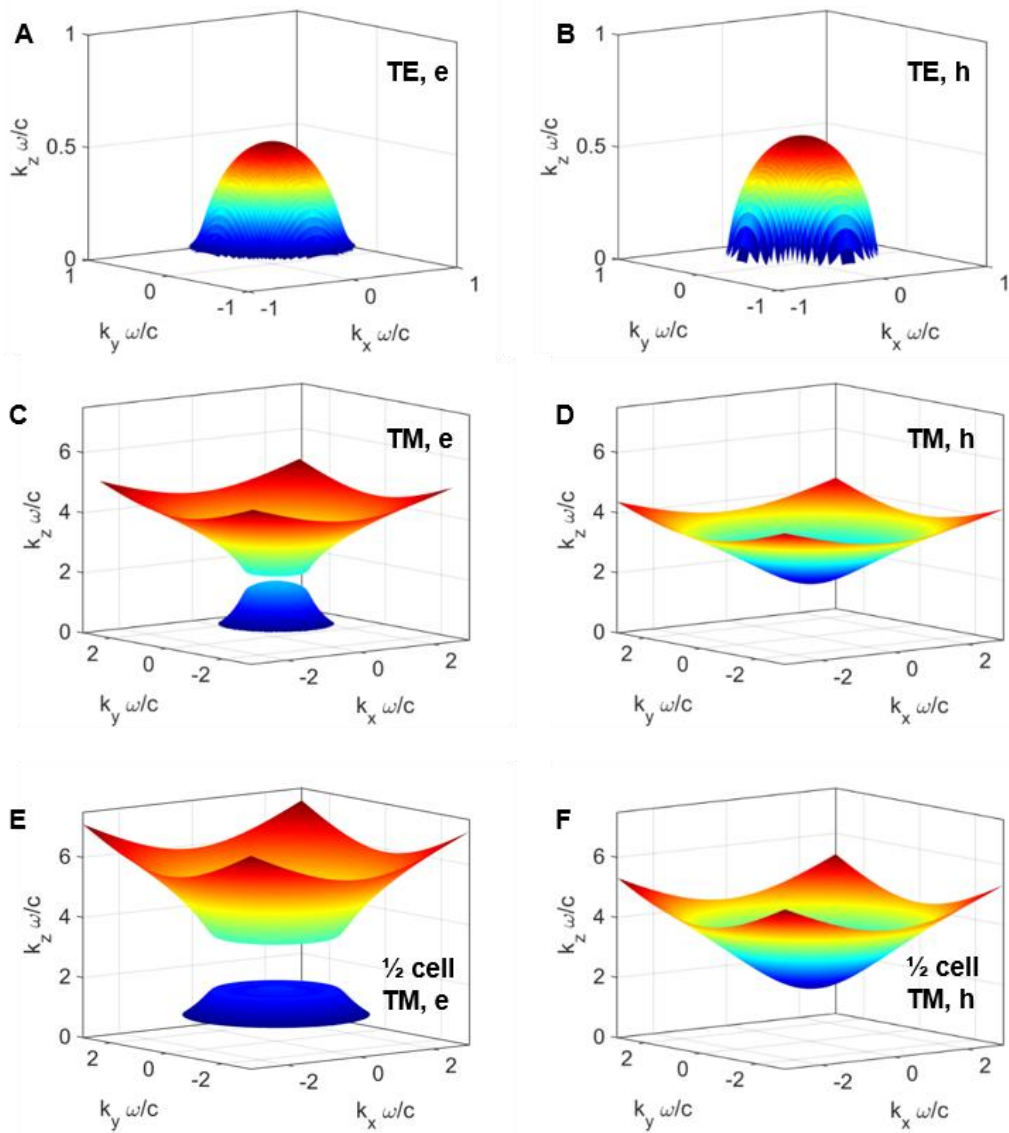

Figure 2. Dispersion of the modes supported by the nanorod metamaterial calculated using the *nonlocal* effective medium theory: (A,C,E) elliptic (denoted e) regime ($\lambda = 550$) nm and (B,D,F) hyperbolic (denoted h) regime ($\lambda = 650$ nm) in the case of (A,B) TE- and (C-F) TM-modes for (A-D) the metamaterial considered in experiments (as in Fig. 1 a) and (E,F) metamaterial with 50% unit cell and nanorod diameter (25 nm diameter, 50 nm period). The scaling does not change the local effective medium parameters of the composite (Figs. 1 C-F).

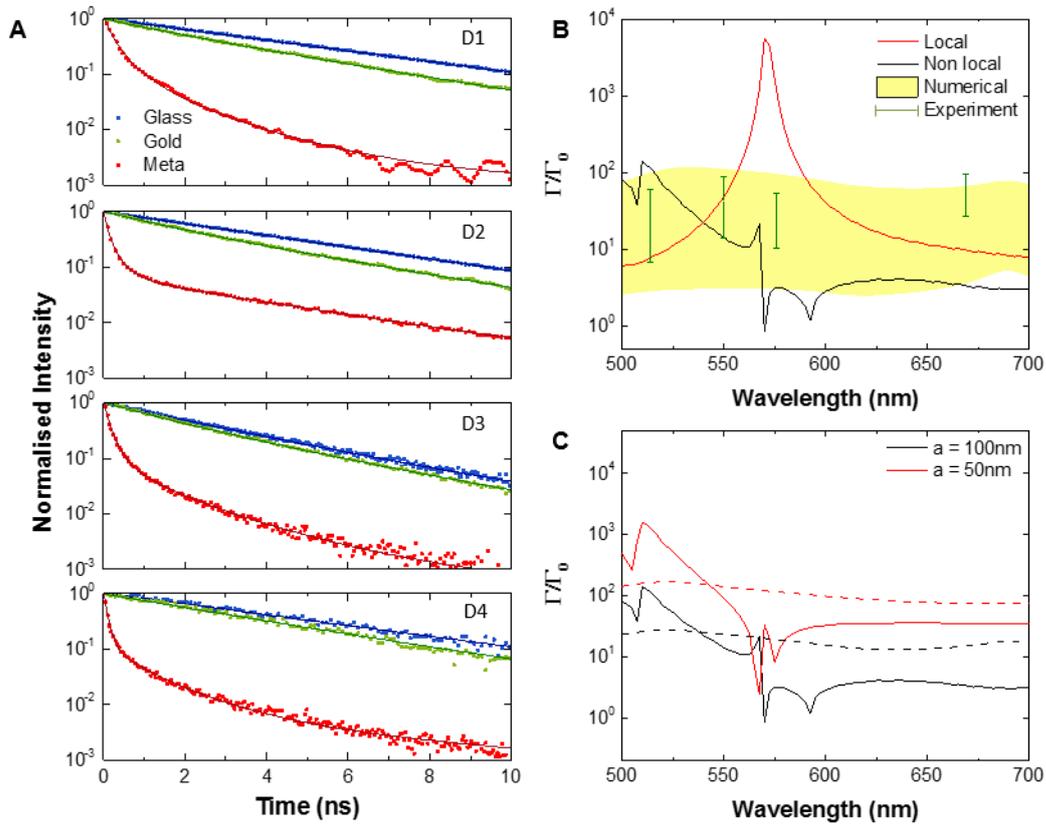

Figure 3. (A) Fluorescence dynamics of the emitters with the emission wavelength in different dispersion regimes of the metamaterial: (blue) on a glass surface, (green) on an Au film, (red) inside the metamaterial. The studied emitters are D1 (fluorescein), D2 (Alexa 514), D3 (ATTO 550) and D4 (ATTO 647N) with the emission wavelengths 514 nm, 550 nm, 575 nm and 670 nm, respectively. Solid lines present dynamics recovered by applying an inverse Laplace transform (see SI) to the experimental data (dots). (B) Spectral dependence of the lifetime averaged over the dipole orientation: (red) local theory, (black) nonlocal theory, (bars) experimental data corresponding to the width of the lifetime distribution at 10% of the modal amplitude (Fig. S3), (shaded area) the width of the lifetime distribution at 10% of the modal amplitude obtained applying the inverse Laplace transform to the decay curves obtained from the averaging over the dipole position within the elementary cell of the metamaterial (Fig. S8), (C) Spectral dependence of the Purcell factor obtained with the nonlocal theory (solid lines) and the full wave numerical modelling at one position within the metamaterial (dashed lines, position 3 in Fig. S8a): (black) the metamaterial as in (B) used in the experiments ($d = 50$ nm; $a = 100$ nm) and (red) the metamaterial with 50% unit cell ($d = 25$ nm; $a = 50$ nm) that yields an identical local effective medium response.

# Supplementary Material.

**Methods.**

*Metamaterial fabrication.* Plasmonic nanorod metamaterials were fabricated by Au electrodeposition into highly ordered nanoporous anodic alumina oxide (AAO) templates on glass cover slips. An Al film of 700 nm thickness was deposited on a substrate by magnetron sputtering. The substrate comprised of a glass cover slip with a 10 nm thick adhesive layer of tantalum pentoxide and a 7 nm thick Au film acting as a weakly conducting layer. Highly-ordered, nanoporous AAO was synthesized by a two step-anodization in 0.3 M oxalic acid at 40 V. After an initial anodization process, the porous layer formed was partly removed by etching in a solution of $H_3PO_4$ (3.5%) and $CrO_3$ (20 $gl^{-1}$) at 70°C, which resulted in anordered, patterned. Then the sample was anodized again under the same conditions as in the first step. The anodized AAO was subsequently etched in 30 mM NaOH to achieve pore widening and remove the barrier layer. Gold electrodeposition was performed with a three electrode system using a non-cyanide solution. The length of nanorods was controlled by the electrodeposition time. Free standing gold nanorod metamaterials were obtained after dissolving the nanoporous alumina template in a mix solution of 0.3 M NaOH and 99.50% ethanol. The nanorod array parameters used in this work are 50±2 nm diameter, 100 nm period and 250±5 nm height (inset in Fig. 1(a)). The extinction spectra of the metamaterial for various illumination angles are shown in Fig. S1.

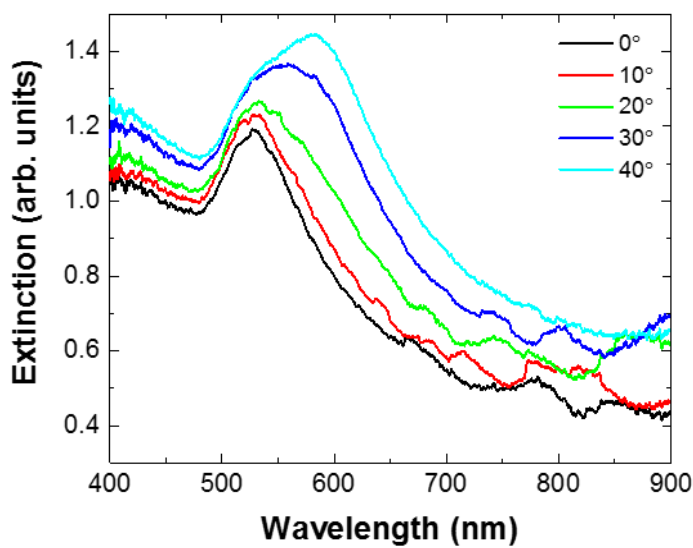

*Figure S1. Extinction spectra of the metamaterials consisting of free standing nanorods immersed in ethanol for different angles of incidence of p-polarised light.*

*Fluorescence lifetime measurements*

Four dye molecules: D1 (fluorescein), D2 (Alexa 514), D3 (ATTO 550) and D4 (ATTO 647N) with the emission wavelengths 514 nm, 550 nm, 575 nm and 670 nm, respectively, covering elliptic, epsilon-near zero, and hyperbolic spectral ranges of the metamaterial were used. For D1 and D2 dyes, the solvent was ethanol and the concentration was $2.5 \times 10^{-5}$ mol/L. For D3 and D4 dyes, the solvent was made of 10mM of Tris-HCL, 1mM of EDTA, 30mM of NaCl and NaOH to adjust the pH of the solution to 7.85, and the concentration of $10^{-6}$ mol/L was chosen. The dye solutions were introduced in the metamaterial using a flow cell, which was positioned on the confocal microscope for fluorescence measurements.

Time-resolved photoluminescence analysis was performed using time-correlated single photon counting (TCSPC) [23] based on a SPC-150 (Becker-Hickl) system. The wavelength of the TM-polarised excitation light from the Fianium supercontinuum laser (< 10 ps pulse duration, 20 MHz repetition rate) was selected (525 nm for ATTO 550, 633 nm for ATTO 647N, 470 nm for Alexa 514 and 470 nm for fluorescein) and focused on the sample with a 100x, NA = 1.49 oil-immersion objective and the resulting photoluminescence (PL) collected by the same objective. Various filters were used to remove the laser contribution to the measured light. Measurements have been taken on the metamaterial samples, thin (50 nm) gold films and reference samples of dyes on glass substrates at the maximum wavelength in the observed spectra. The PL spectra of the different dyes on glass and inside the metamaterial are shown in Fig. S2.

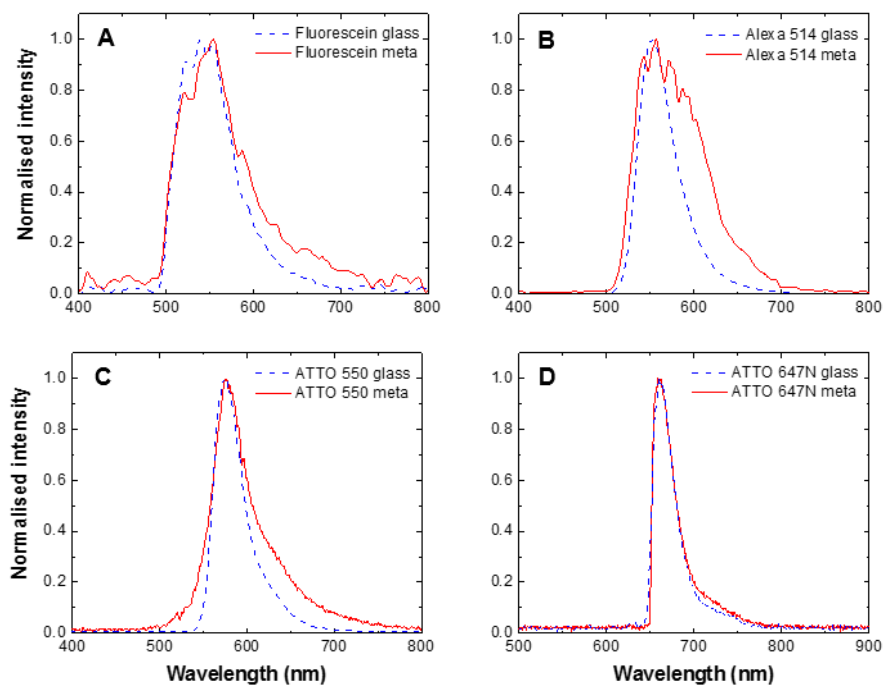

*Figure S2. Emission spectra measured on glass (blue) and inside the metamaterial (red): (A) Fluorescein, (B) Alexa 514, (C) ATTO 550, and (D) ATTO 647N.*

The decay dynamics of the fluorophores on gold and glass substrates is presented in Fig. 3. While all the fluorophores situated in a homogeneous environment show single exponential decay behavior, the presence of surfaces and nanostructures makes the PL dynamics more complex. It should be emphasized, that the effect does not emerge from the strong coupling regime of interaction [17], but is the direct result of the signal collection from the many molecules situated in the probed volume of the sample. Consequently, the different molecules differently positioned and oriented with respect to the Au gold film or nanorods of the metamaterials contribute to the measured signal, all having position and orientation dependent decay rates. While various techniques for fitting the intensity decay exist, we have used the inverse Laplace transform [18] which does not rely on any preliminary assumption and gives the distribution of the measured lifetimes.

*Fluorescence lifetime data analysis.*

*Inverse Laplace transform for evaluation of lifetime distribution*

Lifetime distributions were obtained by solving the following integral equation:

$$I(t) = \int_0^\infty F(s)\, e^{-st}\, ds, \qquad (S1)$$

Where $I(t)$ is the time dependent intensity decay (measured quantity), deconvoluted from instrumental response function (also measured), and $F(s)$ is the relative weight of actual exponential decay components. The deconvolution was performed by applying Tikhonov regularization technique, also introducing low pass filtering, reducing the noise of the decay tail. Since the inverse Laplace transform is known to be an ill-defined problem (especially for analysis of noisy data), an iterative fitting procedure was applied in order to achieve stable results. This type of analysis assumes the exponential decay of the fluorophores that obey the weak coupling regime (no Rabi oscillations) of interaction.

All emitters situated on the glass slide show smooth localized lifetime distributions, peaked in the nanosecond range (Fig. S3). These results agree with the tabulated data, with small deviations related to the type of the solvent that was used [23]. It is worth noting that the glass-ethanol interface has only a small refractive index contrast, and hence could be neglected in the analysis. Next, a plasmonic (Au) interface was introduced for testing the concept of collective lifetime manipulation. The dynamics of fluorophores' lifetimes, modified by the presence of the nearby surfaces, in particular those made of noble metals, is well understood. The key contributors to lifetime modification are mirror-reflected waves, excitation of surface waves and quenching [19]. Using this theoretical formulation and the experimentally measured lifetime distribution of dyes on a glass slide, the theoretical prediction of the distribution on the gold film was obtained. An almost perfect fit between the theoretical prediction taking into account spatial averaging of the emitter position with respect to the metal film, and experimental data suggest the validity of the approach for studies of more complex nanostructures. The spatial average takes into account a random distribution of positions and orientations of florescent molecules in the solution. Taking into account the obtained good correspondence between the experiment and the model, contributions of various possible collective fluorophore concentration-dependent effects, e.g. super-fluorescence[20], [21], can then be ruled out.

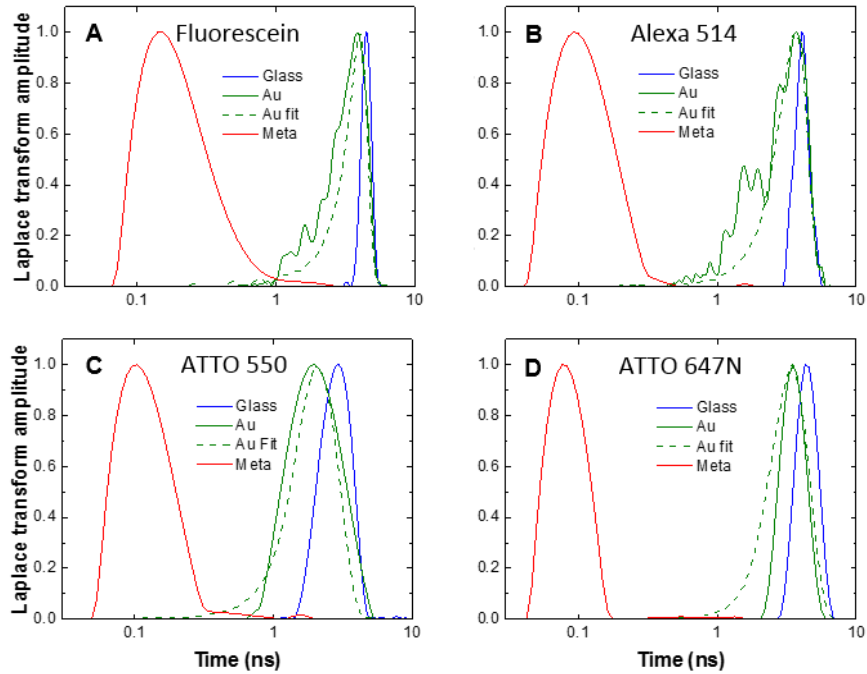

*Figure S3. Experimental fluorescence lifetime distribution of the emitters in different environments extracted using a Laplace transform method from the time resolved PL measurements: (blue) glass substrate, (red) Au film, and (green) inside the metamaterial. Dashed black line represent the lifetime distribution above the smooth Au surface recalculated from the measured lifetime distributions in the dielectric environment shown with blue lines. (A) Fluorescein, (B) Alexa 514, (C) ATTO 550, and (D) ATTO 647N.*

***Fitting collective Purcell enhancement on gold film with analytical formulation***

A semi-analytical expression for Purcell enhancement of a point-like emitter next to an interface between two dielectric materials is well known [19]. It relies on the knowledge of electromagnetic Green's function and its classical quantum correspondence to the spontaneous emission rate (proportionality to the imaginary part of the Green's function)[24]. Hence, given the lifetime distributions of fluorophores on a glass slide, the corresponding distribution on a gold film could be derived, by performing the following integral average:

$$F_{AU}^{Theor}(s) = \int_{d_{min}}^{d_{max}} [P(z) F_{Glass}^{Exp}(s \cdot P(z))] \Psi(z) dz, \quad (S2)$$

where $P(z)$ is the position dependent angular averaged Purcell factor, $F_{Glass}^{Exp}$ is the experimentally measured lifetime distribution of fluorophores on a glass slide, and $F_{AU}^{Theor}$ is the resulting estimate of the lifetime spread on the gold film. $\Psi(z)$ is the distribution density of the fluorophores along the focal depth, and it was assumed to be uniform. The depth of focus $(d_{max} - d_{min})$ was taken to be ~200 nm for the best fitting of the experimental data. The comparison between experimental data for gold and the above theoretical fitting procedure is presented in Fig. S3 and shows excellent agreement for all dyes. The overall peak to peak Purcell enhancement is less than 2. Slight deviations between the fitting and experiment are attributed to noise and to the assumption on the $\Psi(z)$, which in principle, could encapsulate nonuniform pumping intensity of the dyes due to e.g. the reflections from interfaces as well as stochastic inhomogeneity of the solution.

*Analytical modelling*

**a. Effective medium models and modal dispersion.**

In analytical calculations, a layer of plasmonic nanowire metamaterial was represented as a homogeneous layer described by either local [25] or nonlocal [16] electromagnetic theories.

Explicitly, in local effective medium theory (EMT), the permittivity of the metamaterial is given by the Maxwell-Garnett theory:

$$\epsilon_\perp^{mg} = \epsilon_h \frac{(1+p)\epsilon_{Au} + (1-p)\epsilon_h}{(1+p)\epsilon_h + (1-p)\epsilon_{Au}}; \quad \epsilon_{zz}^{mg} = p\epsilon_{Au} + (1-p)\epsilon_h$$

with $\epsilon_h$ and $\epsilon_{Au}$ being permittivities of the host material and gold, respectively. A Drude model, with corrections for restricted mean free path of electrons [26], previously shown to be in quantitative agreement with experimental results [7,15], was used to describe the permittivity of gold. Annealing of metamaterials modifies the restricted mean free path, in turn controlling gold permittivity. This process typically affects the imaginary part of the permittivity somewhat more strongly than it affects the real part. Fig. S4 illustrates the effect of the modification of losses in gold on the effective medium parameters of nanorod metamaterials.

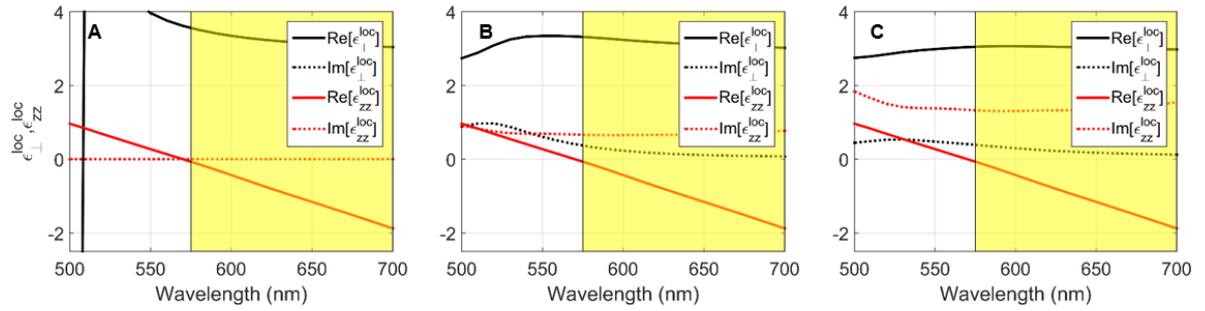

*Figure S4. Local effective medium parameters for nanowire metamaterials with the same geometry $a = 100 nm; r = 25 nm$ but with different absorption in gold, (A) lossless gold, (B) losses as used in the main text, (C) losses twice as large as those considered for (B).*

In nonlocal EMT, the components of the permittivity tensor perpendicular to the optical axis are still described by the Maxwell-Garnett theory ($\epsilon_{xx} = \epsilon_{yy} = \epsilon_\perp^{MG}$), while the component of the permittivity tensor along the optical axis becomes explicitly dependent on the wavevector,

$$\epsilon_{zz}(k_z) = \xi \left( k_z^2 \frac{c^2}{\omega^2} - n_z^{l\,2} \right); \quad \xi = p \frac{\epsilon_{Au} + \epsilon_h}{\epsilon_h - \left(n_\infty^l\right)^2}$$

where $n_z^l$ is the effective refractive index of the cylindrical surface plasmons that propagate in a nanorod composite with the nanorod permittivity $\epsilon_{Au}$ and $n_\infty^l$ represents the limit of $n_z^l$ for perfectly conducting nanorods [16]. These parameters can be calculated either numerically or by solving an eigenvalue-type problem, as shown in [16]. Dispersion of the two TM-polarized modes, supported by the nonlocal metamaterial is given by

$$\left(k_z^2 - n_z^{l2}\frac{\omega^2}{c^2}\right)\left(k_z^2 - \epsilon_\perp^{mg}\frac{\omega^2}{c^2}\right) = -\frac{\epsilon_\perp^{mg}}{\xi}\frac{\omega^2}{c^2}k_x^2$$

The transfer matrix method, described in detail in previous works (e.g. [16]), was used to calculate transmission and reflection through a planar slab of the metamaterial for a given excitation frequency and angle of incidence. Figure S5 shows transmission through the slab of metamaterial used in our work, calculated with the local and nonlocal permittivity models. It can be seen that the two techniques provide almost identical results. Nevertheless, there is a drastic disagreement between predictions of the two techniques when it comes to calculation of emission of the dipole embedded inside the metamaterial.

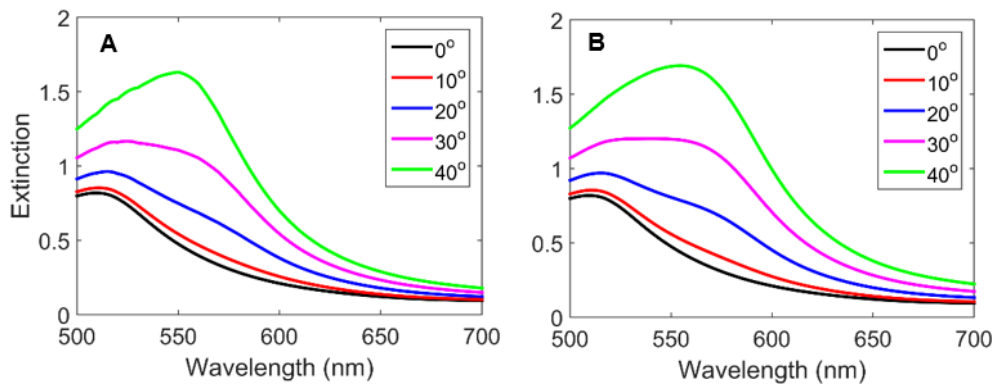

*Figure S5. Extinction spectrum of the nanowire metamaterial submerged in ethanol calculated using local (A) and nonlocal (B) effective medium theories.*

To gain insight into the origin of this disagreement, the dispersion of the modes supported by the nanorod composite, predicted by local and nonlocal effective medium theories was analyzed. It was observed that dispersion of TE polarized modes does not significantly depend on materials absorption or unit cell size. The typical topology of these modes are shown in Fig. 1.

On the other hand, when the metamaterial operates in the elliptical regime, absorption, along with unit cell configuration, plays an important role in defining the topology of TM-polarized modes. This behavior is summarized in Fig. S6.

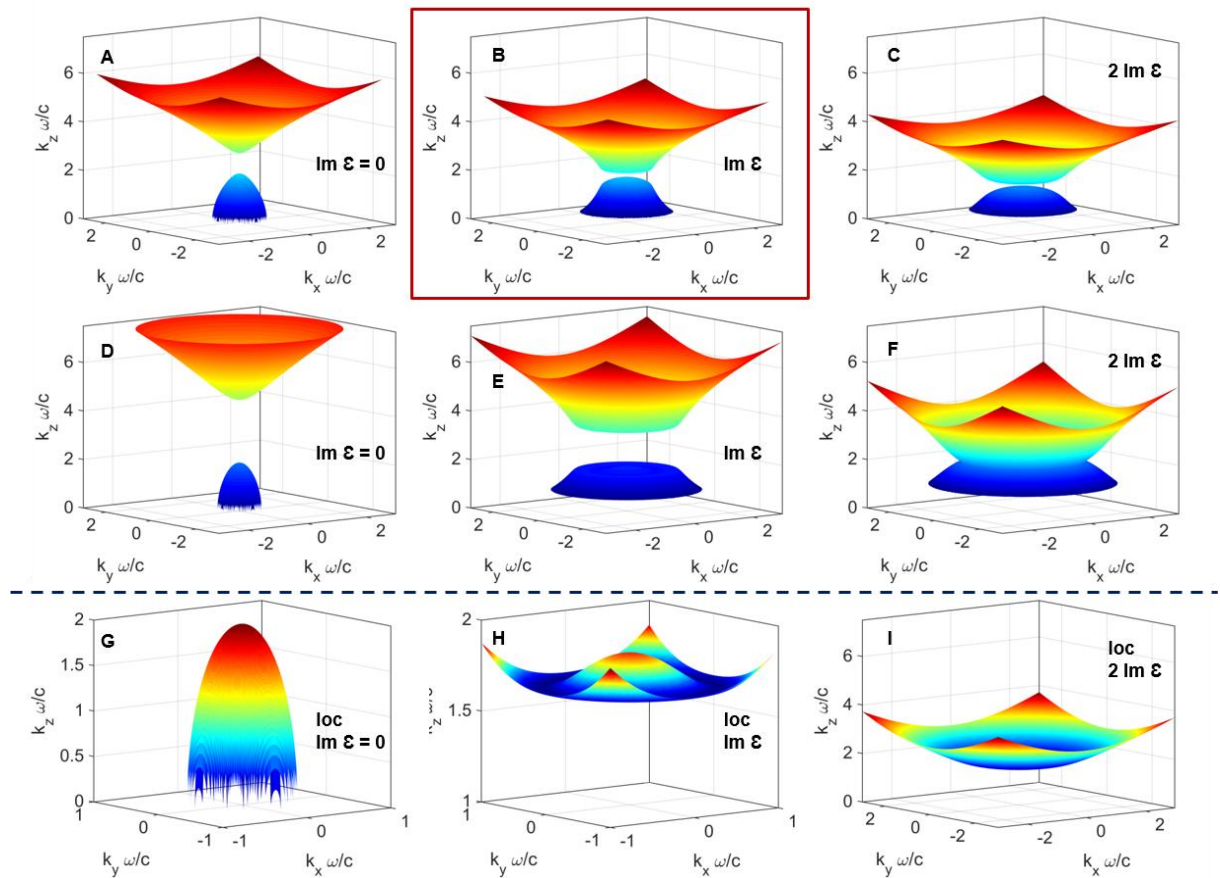

Figure S6. Dispersion of TM-polarized modes supported by nanorod composites in the elliptical regime ($\lambda = 550\ nm$) for different material absorption and geometry: (A,D,G) hypothetical loss-less metamaterial, (B,E,H) absorption as in the experiment, (C,F,I) twice larger absorption than in (B,E,H); (A-F) nonlocal effective medium theory for metamaterials with (A-C) $a = 100\ nm, r = 25\ nm$ as in the experiment and (D-F) 50% unit cell ($a = 50\ nm, r = 12.5\ nm$), (G-I) local effective medium theory; the solid box (B) highlights the dispersion of the modes in the metamaterial used in the experiment; the dashed line separates predictions of nonlocal and local EMTs from each other.

When losses are small, the two TM polarized modes represent well-separated branches in the wavevector space, with the lower branch having elliptical behavior and the upper branch exhibiting hyperbolic-like dynamics (Fig. S6 a,d). As the unit cell becomes smaller, the behavior of the elliptic branch approaches predictions of the local effective medium theory (Fig. S6 g), while the upper branch moves "up", increasing the effective index of the mode. As losses increase, the two TM-polarized modes approach each other (Fig. S6 b,c). Scaling the size of the unit cell of metamaterial still drastically affects the topology of these waves (Fig. S6 e,f). Note that when losses are high, the local EMT converges to the upper (hyperbolic) wave (Fig. S6i); however, in the regime of moderate losses, realized in our work, the local EMT becomes invalid, predicting the response in-between the two real modes of the metamaterial (Fig. S6 h).

When the metamaterial operates in the hyperbolic regime, the elliptical branch of the TM-dispersion cuts off, and the propagation of light is dominated by the hyperbolic branch (Fig. S7). The effect of the spatial dispersion is now essentially limited to a quantitative correction of the dispersion. Decreasing the unit cell size or reducing losses causes the dispersion to approach a "perfect" hyperbola, while decreasing the unit cell alone causes the response to approach predictions of the local effective medium theory.

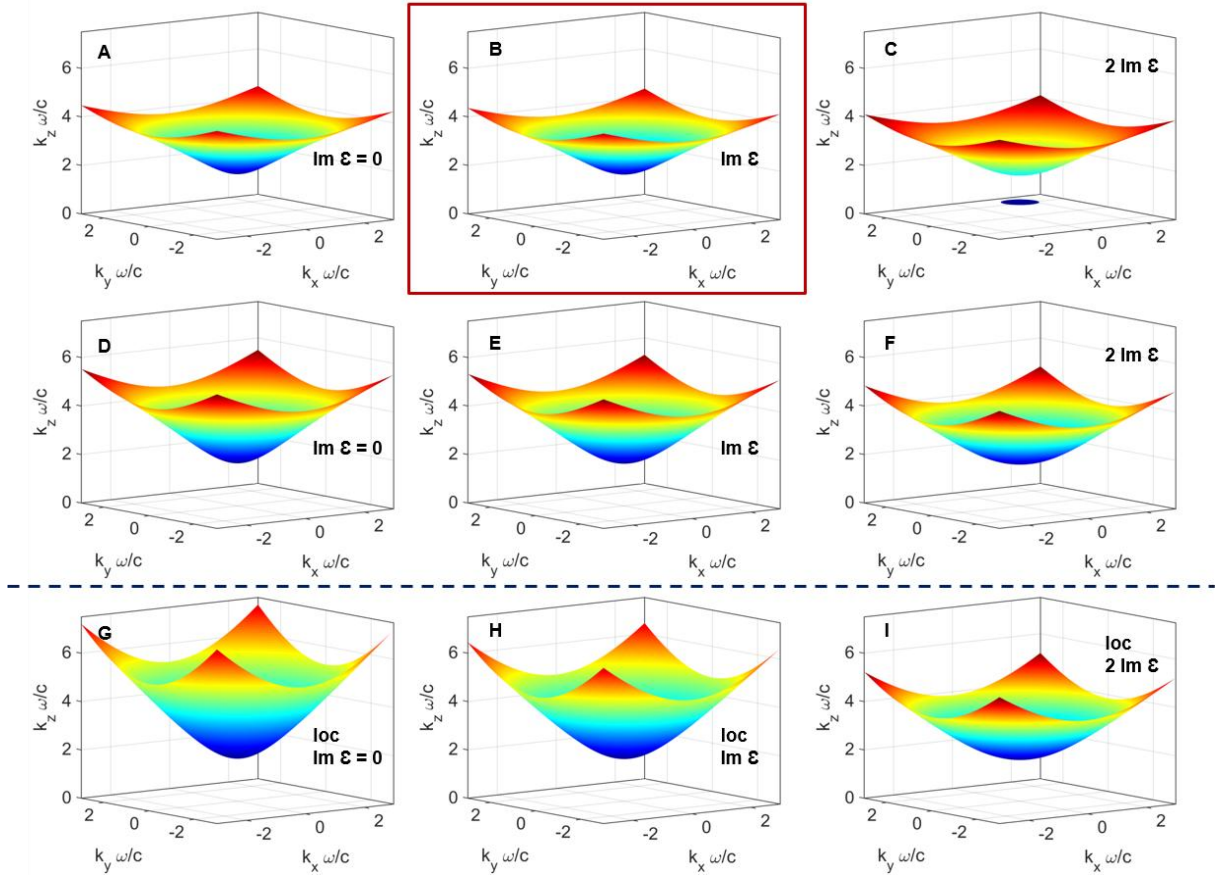

*Figure. S7. Dispersion of TM-polarized modes supported by nanorod composites in the hyperbolic regime ($\lambda = 650\ nm$) for different material absorption and geometry: (A,D,G) hypothetical loss-less metamaterial, (B,E,H) absorption as in the experiment, (C,F,I) twice larger absorption than in (B,E,H); (A-F) nonlocal effective medium theory for metamaterials with (A-C) $a = 100\ nm, r = 25\ nm$ as in the experiment and (D-F) twice smaller unit cell ($a = 50\ nm, r = 12.5\ nm$), (G-I) local effective medium theory; solid box highlights the dispersion of the mode in the metamaterial used in experiment; dashed line separates predictions of nonlocal and local EMTs from each other.*

**b. Fluorescence decay rate enhancement calculations**

A Green's function formalism was used to analyze the emission decay rate enhancement. In this approach, the emission rate enhancement is given by the regular part of the Green's function representing the electric field $\vec{E}$ generated by the point dipole $\vec{d}$ at the location of the dipole [8,11,28]

$$\frac{\Gamma}{\Gamma_0} \simeq \frac{3}{2} \frac{Im(\vec{E}\cdot\vec{d})}{\omega^3 |P|^2}$$

$\hat{z}$- polarized and in-plane polarized dipoles were considered separately, and the total emission rate was calculated as the weighted-average of the two dipole directions.

A transfer-matrix formalism was used to take into account the effect of multiple reflections of the emission by a dipole inside the multilayered composite [27]. In order to avoid singularities caused by placing the dipole inside a lossy medium [26], the dipole was placed inside a 1 nm thin layer of lossless medium, cut out inside the conventional lossy metamaterials. In nonlocal calculations, emission into two TM-polarized modes were considered as independent emission channels; however, our calculations suggest that the hyperbolic-like mode dominates emission in both elliptic and hyperbolic regimes, similar to what has been previously reported in emission inside infinite idealized metamaterial [28].

*Numerical modelling*

Numerical modeling was performed using the finite element method (FEM) implemented in Comsol Multiphysics software. The Purcell factor at various positions inside the nanorod metamaterial was calculated as a ratio of power flow from a point dipole placed there and the corresponding value for a dipole in the uniform dielectric. Two methods were used for the power flow estimation: 1) via a Poynting vector flux through a small sphere (5 nm radius) enclosing the point dipole and 2) via energy dissipation rate of the dipole, $W(\mathbf{r}) = -\frac{1}{2}\text{Re}[\mathbf{E}^*(\mathbf{r})\cdot\mathbf{J}(\mathbf{r})]$ (where $\mathbf{E}(\mathbf{r})$ is the electric field produced by the dipole at the point of its location and $\mathbf{J}(\mathbf{r})$ is the dipole current). Both methods showed excellent agreement. The arbitrary orientation of the dipole was taken into account by averaging the Purcell factor over all dipole orientations, which in this case presents a linear combination of the corresponding values for emitters with a dipole moment along the three coordinate axes $P(\mathbf{r}) = \frac{1}{3}P_x(\mathbf{r}) + \frac{1}{3}P_y(\mathbf{r}) + \frac{1}{3}P_z(\mathbf{r})$ [29]. The results of these classical electromagnetic simulations can be directly mapped onto the quantum decays of emitters, by employing the classcial quantum correspondence of radiation reaction forces [30]. For the numerical simulations, in order to simulate infinite number of nanorods in the arrays, we gradually enlarge the number of rods in the finite-size arrays with periodic boundary conditions on their sides. The convergence of the Purcell factor with the number of the nanorods within the simulation domain (Fig. S8) confirms that a 10x10 nanorod array with periodic boundary conditions can be used to simulate the behaviour of an infinite metamamterial. The averaging over the dipole position within the primitive cell of the array of the metamaterial was performed, assuming a uniform distribution of the emitters with a position dependent decay rate and local excitation efficiency, by a pump light illuminating the metamaterial from the substrate side: $W(t) \propto \int_{e-c} |\mathbf{E}_{\text{pump}}(\mathbf{r})|^2 Q_{\text{loc}}(\mathbf{r}) n(\mathbf{r}) \exp(-P(\mathbf{r}) \cdot \Gamma_0 t) d^3\mathbf{r}$, where the integration is done over the dye-filled volume within the elementary cell, taking account the distance dependent number of emitters $n(\mathbf{r})$. $Q_{\text{loc}}(\mathbf{r})$ is a local quantum efficiency defining how much of the emitted power is measured in the far-field (the internal quantum yield of the emitters was taken to be 1 in all the simulations). The local excittaion intensity was found to vary by only 8% between 470 and 633 nm excitation wavelength used in the experiment and were averaged for these values. Since this equation provides a multiexponential decay, a procedure equivalent to the above described Laplace transform was applied in order to recover the lifetime distribution.

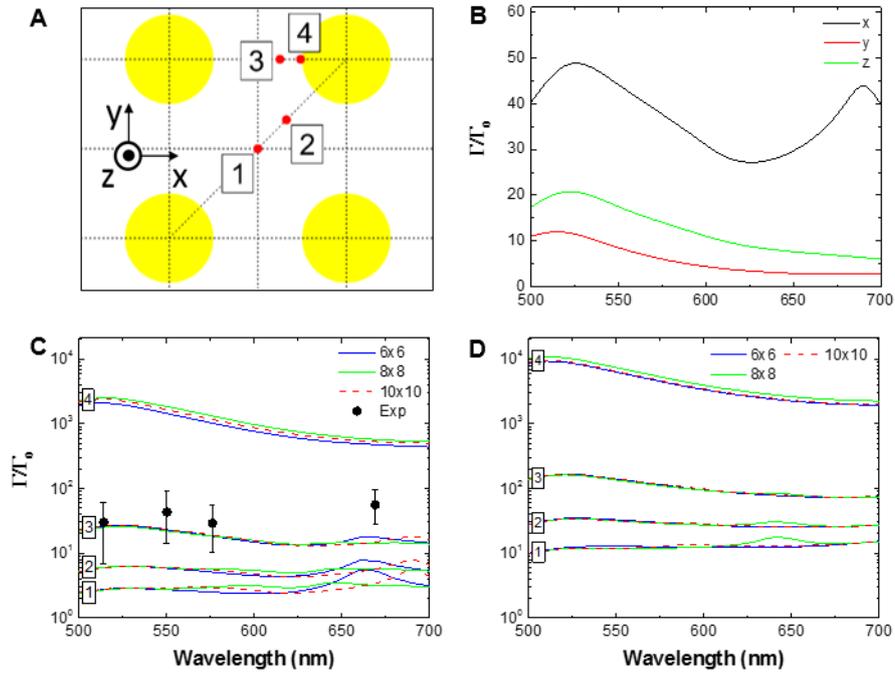

*Fig S8. Numerical simulations of the Purcell factor in metamaterials. (A) Schematic of the metamaterial with the position of the emitters used in the simulations. All the emitters are situated in the middle of the nanorod length. (B,C,D) Spectral dependence of the Purcell factor for (B) a dipole with different orientations at position 3 inside the metamaterial, (C,D) a randomly oriented dipole at different positions inside the nanorod metamaterial (as indicated in (A)) for the metamaterials with (B.C) period $a = 100$ $nm$ and nanorod radius r = 25 nm (as in the experiment) and (D) a = 50 nm and r = 12.5 nm, corresponding to the same local effective medium parameters. The colored lines correspond to different sizes of the finite nanorod array used in the simulations (as indicated in the legends) showing the convergence to the behavior of the infinite metamaterial; symbols in (c) represent the experimental data with points indicating the maximum of the lifetime distributions (Fig. S3) and the bar corresponding to the width of the distribution at 0.1 amplitude. In all simulations the internal quantum yield of the emitter was considered to be 1.*